# Characteristics of light reflected from a dense ionization wave with a tunable velocity

A. Zhidkov[1], T. Esirkepov[2], T. Fujii[1], K. Nemoto[1], J. Koga[2], S.V. Bulanov[2]

[1] Central Research Institute of Electric Power Industry, 2-6-1 Nagasaka, Yokosuka, Kanagawa, 240-0196 Japan
[2] Kansai Photon Science Institute, Japan Atomic Energy Agency, 8-1 Umemidai, Kizugawa, Kyoto, 619-0215 Japan

**Abstract.** The optical field ionization of a transparent media by two, cylindrically focused femtosecond laser pulses may result in production of an ionization wave (*IW*). Velocity of such a quasi-plane *IW* in the vicinity of pulse intersection can be tuned by changing the intersection angle and can even exceed the speed of light. We study the conversion of a coherent light to x-rays by means of particle-in-cell simulation and by solution of continuous equation with the correct current:

$$j(x,t) = -e \int (dN_e/dt_0) v(t,t_0) dt_0$$

. X-ray spectrum of converted lower frequency light changes from the monochromatic to a high order harmonic-like with the duration of ionizing pulses. The conversion efficiency can be increased via suppression the energy of the generated magnetic field

## INTRODUCTION

Recently, the interest in relativistic or flying mirrors, which are the relativistic objects that can reflect radiation coherently shortening the radiation wavelength [1], has grown [2] with developing of multi-terra-Watt femtosecond lasers [3]. Serial theoretical and experimental works have been dedicated to the interaction of coherent electro-magnetic radiation with relativistic objects [4-6,7]. The main goal of the researches is shifted now to searching compact coherent x-ray sources [8,9] with use of plasma. In these researches the ionization waves look very attractive because, in contrast to electron beams and plasma waves, there are no energetic particles in the scheme and the optical density of *IW*s does not depend on their velocity; there is no energy transmission from the mirror to x-rays, therefore, high energy ionizers are not necessary.

Although the theory of light reflection from *IW* has been presented [6, 7], its elaboration and developing are important for understanding the characteristics of the *IW* mirrors produced by femtosecond laser pulses. In Ref. [10] the ionization wave in *Ar* produced by an intense fs laser pulse has been proposed as a mirror for $CO_2$ laser pulses. However, the relativistic regime with $V_{gr}$ close to the speed of light was achieved only by increasing the intensity of the ionizer which drastically reduces the reflectivity of that mirror. Here, we numerically study characteristics of *IW* mirrors generated by crossing laser pulses in high pressure gas and their scattering of a longer wavelength radiation. The conversion of laser light is always 100%. However, a considerable portion of the energy remains in plasma as the static magnetic field.

## IONIZATION WAVE' CONVERTERS

We consider a new scheme for *IW*s with tunable velocities that can even exceed the speed of light. Such *IW*s can be generated by the optical field ionization of a transparent media as shown in Fig.1. The resulting *IW* has the velocity equal to $V_{gr}/\sin(\theta/2)$, where $V_{gr}$ is the velocity of a single ionization waves and $\theta$ is the angle between the ionizers. When $\theta$ is less than $180°$ the velocity of *IW* becomes larger than $V_{gr}$.

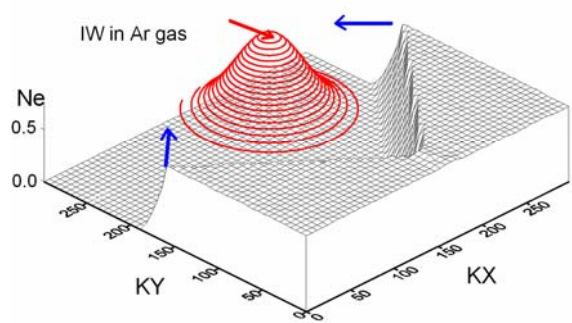

**FIGURE 1.** Calculated density of an *IW* in 3 atm Ar gas produced by two Ti-Sph laser pulses of 30 fs duration and $a_0$=0.5 moving with the group velocity $V_{gr}$=0.7$c$ and crossing at $\theta$=90° and the scheme of the light scattering. The blue arrows show the direction of ionizing pulse propagation, the red one shows the propagation of a probing light.

[For $V_{gr}$=0.7$c$ the angle must be 90° for the mirror velocity to become equal the speed of light.] The use of different frequencies for ionizing and for scattering pulses may make the process very efficient, close to 100%: the optical density of the *IW* plasma does not depend on the *IW* velocity. Shaping of focusing mirrors allows a velocity modulation of the combined ionization wave. Even though a considerable portion of laser light scattered by *IW* remains in plasma in the form of magnetic field, efficiency of such an *x*-ray converter may exceed that of all known *x*-ray sources.

The typical density distribution in the ionization wave in *Ar* gas is also shown in Fig.1. To find it we solve the balance equation

$$dN_{z+1}(\mathbf{r},t)/dt = \alpha(z, E/E_{at})N_z(\mathbf{r},t), \quad (1)$$

where $\alpha$ is the optical field ionization (OFI) rate of an ion with charge $z$ at the laser field strength $E$; $E_{at}$ the atomic field, the electron density $N_e = \sum z N_z$ [10] in a square area irradiated by two Ti-Sph laser pulses of Gaussian-shape with $a_0 = eE/mc\omega = 0.5$ [$\omega$ the laser frequency] and duration 30 fs that move with $V_{gr}$=0.7$c$ and cross at $\theta$=90°. The pulses produce an ionization wave moving with $V=c$ with density ramp of $L<10$ μm and quite planar, ~40 μm in the transverse direction. With the pulse duration of 8 fs at the same $a_0$, the density ramp becomes shorter, down to 2 μm.

Actual density distribution in *IW*s is more complicated. The electric field in the laser pulse oscillates as $E \sim E_0 \cos(\omega t)$, while the rate $\alpha$ is proportional to $\exp(-A/|E|)$ with $A$ is the constant depending on the ion characteristics. As a result, the solution of Eq. (1) has a ladder-like shape with the stair length order of the ionizer wavelength.

We consider the scattering of laser pulses by both a sharp and a realistic density ramp ionization wave. The calculations are performed by solving one-dimensional (*1D*) fluid equations and by using a two dimensional (*2D*) particle-in-cell method in the laboratory reference frame to avoid the problem with the conversion of the complex initial conditions at mirror's velocities exceeding the speed of light. In the *1D* approach, the standard set of equations:

$$\frac{\partial H_Z}{\partial x} = -\frac{\partial E_Y}{c\partial t} + \frac{4\pi}{c}j_Y;$$
$$\frac{\partial E_Y}{\partial x} = -\frac{\partial H_Z}{c\partial t}; j_Y = -eN_e p_Y/\gamma; \quad (2)$$
$$\frac{dp_Y}{dt} = -e(E_Y - p_X H_Z/\gamma mc);$$
$$\frac{dp_X}{dt} = -ep_Y H_Z/\gamma mc; \gamma = \sqrt{1 + (p_X^2 + p_Y^2)/(mc)^2}$$

with the initial conditions:

$$H_Z = -E_Y = a(t+x/c)e^{i\omega t(t+x/c)};$$
$$a = a_0 e^{-(t+x/c)^2/\tau^2} \quad (3)$$

is solved assuming the continuous electron current with the electron density obeying

$$N_e = N_0 f(X(t) - x)$$

with $f(X)$ the density profile, $X(t)$ is a mirror trajectory, and $N_0$ is the maximum electron density with the density profile produced by 8 and 30 fs ionizing pulses. The current is calculated as

$$J(x,t) = \int_0^t dt_0 u_e(x,t,t_0) dN_e/dt_0, \quad (4)$$

with the electron fluid velocity $u_e(x, t_0, t_0)$=0: the electrons appear with zero velocities. The density ramp has a ladder-like shape with a stair size equal to the half wavelength of ionizing laser pulse. In this case, the Eq. (4) can be presented as a sum.

For the one-step shape *IW* moving with a constant velocity $V$, the set of equations can be reduced to the equation for the vector potential:

$$\left(\frac{\partial^2}{\partial t^2} - c^2 \frac{\partial^2}{\partial x^2}\right) A(x,t) +$$
$$\omega_{pl}^2 \eta(t - x/V)[A(x,t) - A(x, t = x/V)] = 0$$

with $\omega_{pl}$ the plasma frequency and $\eta$ the Heaviside function. It can be solved analytically for $V<c$ upon using the reference frame moving with velocity $V$ with the conventional boundary condition [6]. The scattering of a plane electro-magnetic wave by a rapidly moving density ramp is characterized by five fields: an incident wave, $\omega$, plasma wave, $\omega_{pl}$, two scatted waves, $w_{1R}$ and $w_{2R}$, and a static magnetic field: $k_B = \omega(1/V + 1/c)$ and

$$\frac{\omega_{1,2R}}{\omega} = \frac{1}{1-V/c} \pm \frac{V/c}{1-V/c}\sqrt{1 + \frac{1-V/c}{1+V/c}\frac{N_e}{N_{cr}}}; V<c$$

with the infinite first frequency at $V \to c$ and with the second $\omega_{2R} \to \omega(1 + N_e/2N_{cr})$.

The intensity of the components decreases with $V$ approaching to $c$. In Fig. 2a, b, typical spectra calculated in the *1D* approximation are presented in the case of a low intensity ($a_0$=10$^{-3}$) coherent light scattered by a steep ionization wave moving with different velocity. At $V<c$, spectrum are monochromatic, the conversion efficiency is about 2-5%. The frequencies of x-rays follow Eq. (1). 95% of laser pulse energy remains in plasma in the form of a static magnetic field. However when *V* exceeds the speed of light *c*, the frequency dependency becomes different from those given by Eq. (1) as shown in Fig.2b; the phase matching conditions as in Ref. [6] cannot be applied. One can see that the *x*-ray frequency is much higher than the common Doppler shift: for $V/c=1.1$ the up-shift is 400 that corresponds to $V/c=0.995$.

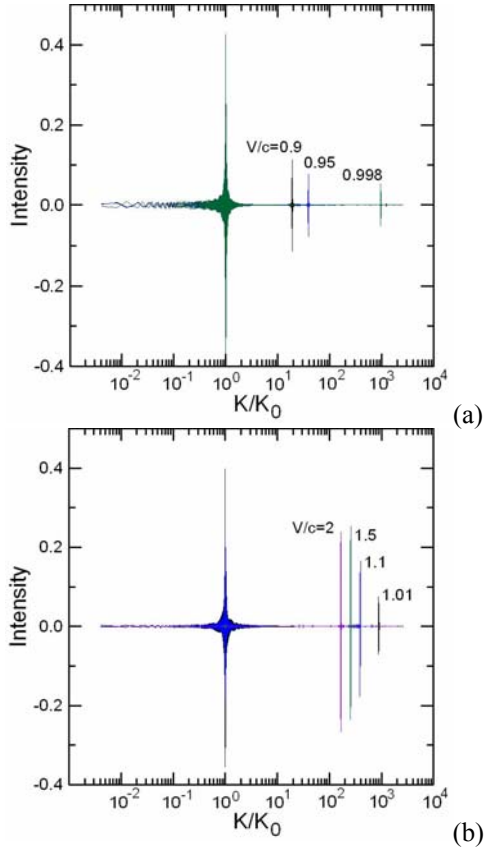

(a)

(b)

**FIGURE 2** The *x*-ray spectra produced via the scattering of a *Ti-Sph* laser pulse with intensity $a_0=10^{-3}$ by a 8fs-*IW* with electron density $N_e=10N_{cr}$ [the emulation of pumping by *KrF* laser pulses] moving with velocity $V<c$ (a) and $V>c$ (b). Data are given at the half-pulse scattering time.

An increase of density ramp length, as in the case of 30 fs-duration ionizing laser pulses, results in appearance of a multiplet spectrum with quite monochromatic components both in the case of $V<c$ and $V>c$ as seen in Fig.3a,b. The number of components is proportional to the number of stairs at the density ramp. The infinite increase of this number results in a broad harmonics-like spectrum. The growth of the intensity of the incident light to $a_0=0.1$ shows different spectra at $V<c$ and $V>c$. In the first case, $V<c$, a monochromatic spectrum of scattered x-rays is generated, while at $V>c$ a broad spectrum is found as seen in Fig.3b. A further increase of the intensity is out of scope of this approach.

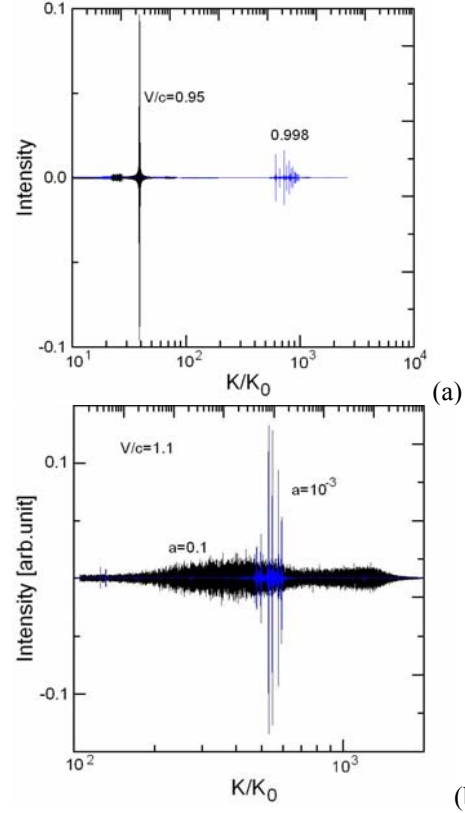

(a)

(b)

**FIGURE 3** The *x*-ray spectra produced via the scattering of a *Ti-Sph* laser pulse by a 30fs-*IW* with electron density $N_e=10N_{cr}$: $a_0=0.1$, $V/c=0.95$ and $V/c=0.998$ (a); $V/c=1.1$ and $a_0=10^{-3}$ mand $a_0=0.1$ (b). Data are given at the half-pulse scattering time.

The temporal behavior of the scattered x-rays is illustrated by Fig. 4 a-d. A strong compression of radiation is observed in the cases $V<c$ as it follows from difference of velocities $|c-V|$; in the case of $V>c$ compression is weaker. The strong magnetic field remaining in plasma is also shown in Fig. 4a,c. It is generated by the transverse constant electron current, $j_0 \sim \cos(\omega x(1/V+1/c))$ and takes a considerable portion of the energy. Ways of lessening of the magnetic field and increasing of the *x*-ray conversion efficiency as proposed in Ref [6] are out of the scope of this letter.

The direct *2D* particle-in-cell simulation is performed for a *IW* with constant $V=0.9c$. Even though the numerical resolution does not allow the calculation for *V* close to the speed of light, the *2D* simulation confirms results of 1*D* simulation giving a good

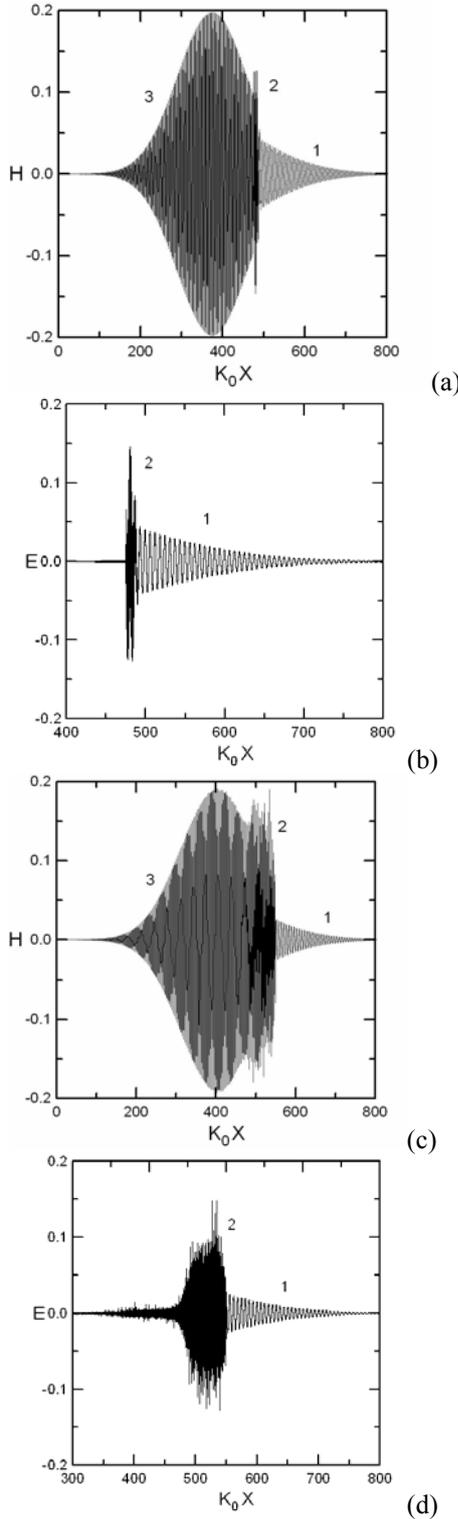

**FIGURE 4** Magnetic field (a),(c) and electric field (b),(d) distribution after complete scattering of 150 fs *Ti-Sph* laser pulse with $a_0=0.1$ at the condition given in Fig.3 at $V=0.9c$ and $V=1.1c$.

agreement for $V=0.9c$ both for the compression and spectrum.

*IW*s can also be velocity-modulated by shaping the focusing mirrors. The most important result of the modulation of the mirror velocity is shown in Fig. 5. One can see the appearance of 4 consequent attosecond *x*-ray pulses with difference in time of 9 fs between them. The closer the mirror speed approaches the speed of light, the stronger pulse compression and the shorter the *x*-ray wavelength. Such modulated *x*-ray may become very useful for the correlative attosecond measurements [8].

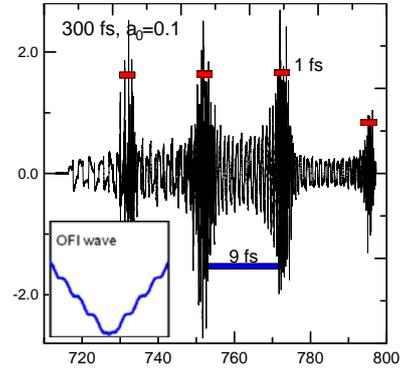

**FIGURE 5** Spatial distribution of electric field after scattering of a 300 fs *Ti-Sph* laser pulse with intensity $a_0=0.1$ by a 30fs-*IW* with electron density $N_e=10N_{cr}$ moving with velocity $V/c=[0.8+0.19\cos(\omega t/20)]$.

The *x*-rays are expected to be coherent. For $V<c$, the typical condition $N_e\lambda^3\gg 1$ transforms for the scattering by *IW*s as $(N_{e0}\gamma)(\lambda_0/2\gamma)^3\gg 1$ where $\gamma=(1-V^2/c^2)^{-1/2}$. The typical wavelength of the *x*-rays can be estimated as $\hbar\omega_{max}=\hbar\omega_0 4\gamma^2$; they remain coherent up to $\hbar\omega_{max}=8\times10^{-5}N_e^{2/3}[cm^{-3}]\lambda[\mu m]$ eV. For the erbium laser with 2.8 μm wavelength and $N_e=10^{21}$ cm$^{-3}$, the maximal energy of the coherent *x*-rays can exceed 2 MeV and their number approaches $10^7$ particles per 1J of scattered laser energy.

We have demonstrated that the optically dense relativistic mirrors (*x*-ray converters) can be produced on the basis of the optical field ionization of transparent media. The use of two or several ionizers allows the tuning of mirror velocity $V<c$ and $V>c$ and, therefore, tuning the frequency of the scattered *x*-rays. Due to the difference in the frequencies of ionizing and scattering laser pulses, the mirrors are optically dense for incident laser pulses resulting in their high reflectivity. Even though the magnetic field generated as a result of scattering takes the most of energy of incident light, the conversion efficiency is practically high.

Upon applying 8 fs laser pulses as ionizers, the *x*-ray spectrum has been shown to be as monochromatic as the incident laser light is. With more practical 20-30 fs duration laser pulses, the spectrum becomes a multiplet with the number of components equal to the number of ionization stairs. At higher intensity of converted radiation the spectrum does not change much at $V<c$ and becomes a harmonics-like in the case of $V>c$. Nevertheless due to the spectrum asymmetry, the efficiency of x-ray conversion is higher at $V>c$. The velocity modulation, easily achieved for such mirrors, may give a set of attosecond x-ray bunches which can be used for ultra-fast measurements. The simple design of the mirrors lets avoid damages.